\documentstyle[12pt]{article}
\def\lsim{\mathrel{\rlap {\raise.5ex\hbox{$ < $}}
{\lower.5ex\hbox{$\sim$}}}}

\newcommand{\pr}{\paragraph{}}
\newcommand{\be}{\bea}
\newcommand{\ee}{\eea}
\newcommand{\bea}{\begin{eqnarray}}
\newcommand{\nn}{\nonumber}
\newcommand{\eea}{\end{eqnarray}}

\newcommand{\nk}{\noindent}

\def\gappeq{\mathrel{\rlap {\raise.5ex\hbox{$>$}}
{\lower.5ex\hbox{$\sim$}}}}

\def\lappeq{\mathrel{\rlap{\raise.5ex\hbox{$<$}}
{\lower.5ex\hbox{$\sim$}}}}

\def\H{H\hskip-8.5pt/\hskip2pt}

\catcode`\@=11 

\def\lsim{\mathrel{\mathpalette\@versim<}}
\def\gsim{\mathrel{\mathpalette\@versim>}}
\def\@versim#1#2{\vcenter{\offinterlineskip
    \ialign{$\m@th#1\hfil##\hfil$\crcr#2\crcr\sim\crcr } }}

\def\t1{{\tilde 1}}

\def\nn{\nonumber}

\def\gappeq{\mathrel{\rlap {\raise.5ex\hbox{$>$}}
{\lower.5ex\hbox{$\sim$}}}}

\def\lappeq{\mathrel{\rlap{\raise.5ex\hbox{$<$}}
{\lower.5ex\hbox{$\sim$}}}}

\begin{document}

\begin{titlepage}

\begin{flushright}
ACT-07/96 \\
CERN-TH/96-143 \\
CTP-TAMU-21/96 \\
OUTP-96-24-P \\
hep-th/9605211 \\
\end{flushright}
\begin{centering}
\vspace{.1in}
{\large {\bf Distance Measurement and Wave Dispersion in
a Liouville-String Approach to
Quantum Gravity}} \\
\vspace{.2in}
{\bf G. Amelino-Camelia$^{a}$},
{\bf John Ellis$^{b}$},
{\bf N.E. Mavromatos$^{a,\diamond}$} \\
and \\
{\bf D.V. Nanopoulos}$^{c}$ \\
\vspace{.03in}
\vspace{.1in}
{\bf Abstract} \\
\vspace{.05in}
\end{centering}
{\small
Within a Liouville approach to non-critical string
theory, we discuss space-time
foam effects on the propagation of low-energy particles.
We find an induced frequency-dependent dispersion
in the propagation of a wave packet,
and observe that this
would affect the outcome of measurements
involving low-energy particles as probes.
In particular, the maximum possible order of magnitude
of the space-time foam effects would give rise to an
error in the measurement of distance comparable to
that independently obtained in some recent heuristic
quantum-gravity analyses.
We also briefly compare these error
estimates with the precision of astrophysical measurements.}

\vspace{0.2in}
\nk $^a$ University of Oxford, Dept. of Physics
(Theoretical Physics),
1 Keble Road, Oxford OX1 3NP, United Kingdom,   \\
$^b$ Theory Division, CERN, CH-1211, Geneva, Switzerland,  \\
$^{c}$ Center for
Theoretical Physics, Dept. of Physics,
Texas A \& M University, College Station, TX 77843-4242, USA
and Astroparticle Physics Group, Houston
Advanced Research Center (HARC), The Mitchell Campus,
Woodlands, TX 77381, USA. \\

\vspace{1.2in}
\begin{flushleft}
May 1996 \\
$^{\diamond}$ P.P.A.R.C. Advanced Fellow \\
\end{flushleft}
\end{titlepage}

\newpage

\section{Introduction}

Two of the most important
problems of Modern Physics are the lack of a consistent
quantum theory of gravity, and a satisfactory description
of the quantum measurement process.
As regards the first of these problems,
so far there is no mathematically consistent
local field theory of gravity
which is compatible with quantum mechanics. Although
the classical limit of gravity, General Relativity,
has been verified experimentally with good accuracy,
profound difficulties arise when one seeks to
combine it with the quantum features of our world.
These difficulties have many aspects. The absence of a
(renormalizable) path integral for quantum geometries
in four space-time dimensions is a very basic formal issue~\cite{book}.
Puzzles and inconsistencies arise even in
certain weak-coupling limits of general relativity with
quantum mechanics. One such
example is the semiclassical quantization
of macroscopic black holes: when one quantizes
the fields of low-energy point particles in such backgrounds,
there appears Hawking radiation due to quantum evaporation
of the black hole~\cite{hawk1}. The latter results in entropy production
and therefore the apparent evolution of pure quantum-mechanical
states to mixed ones.
As the macroscopic black hole evaporates, the energy and
entropy of the radiation field increases.
Obviously, such a situation {\it cannot be} described by treating
the quantum field theory subsystem as {\it closed}.
Indeed, the general view is that
there is information loss across the event horizon surrounding
the black hole.
\pr
This apparent difficulty also presents another problem
associated with the ill-defined nature of the
quantum-gravitational path-integral measure. If
{\it microscopic} Planck-scale
fluctuations of the black hole type exist, which is to be expected
since they exist classically, the possibility arises
then that the associated Hawking radiation will affect
the purity of quantum-mechanical states interacting with
this quantum-gravitational space-time foam~\cite{hawk2,ehns}.
At present there is no known consistent way of formulating
scattering-matrix elements for the propagation of quantum
matter incorporating recoil effects in black hole backgrounds.
And the black hole is but one example
of a topologically non-trivial gravitational fluctuation: there
are likely to be others that could cause similar or worse problems.
\pr
The measurement problem is one of the great
mysteries of quantum theory itself, even if
gravitational interactions are ignored.
Quantum mechanics is a non-deterministic theory,
whereas the physical laws governing
the macroscopic limit of the observable world
have deterministic formulations.
Many physicists have attempted
to shed some light on this question by
exploring various mechanisms
that could lead to the transition
from the microscopic quantum world to the macroscopic classical
world. The most influential idea in the field is that
of the {\it decoherence} of a quantum system in interaction
with a macroscopic measuring apparatus~\cite{zurek,emohn}.
The idea is that, during a measurement process,
the quantum system is an {\it open} system,
and as such it should be described by a density matrix.
After some time $t_D$, characteristic
of the system, the off-diagonal elements
of its density matrix decay {\it exponentially} in time.
The time $t_D$ is inversely proportional to the number of
`atoms' of the measuring device~\cite{emohn}. Hence,
the collapse time for macroscopic systems
is diminished significantly, and the transition
to classicality is achieved.
\pr
Until recently, there were objections~\cite{zurek} to this approach
to classicality based on the argument that the
decoherence (or density matrix ) approach
refers to an ensemble of theories, and hence does
not describe correctly the behaviour of a {\it single} quantum
mechanical system, which should in principle be described
by a state vector. As a counterargument,
we would emphasize that the above approach -
which is more general than the state vector approach -
is the correct one, since
the interaction with the measuring device opens up
the microscopic system. This means that one must
average over possible histories
in the path integral, and hence over theories~\cite{hartle}.
The state-vector picture is inadequate in such a case,
whilst the
density-matrix picture is an appropriate description.
\pr
A mathematically satisfactory approach which formally
reconciles the two pictures has been developed
recently in ref.~\cite{gisin}. It was shown that,
for quantum mechanical systems in interaction with an environment
whose details are irrelevant to a major extent,
a state-vector picture can be written down,
but the time evolution is {\it not} of Schr\"odinger type.
Instead, it resembles {\it stochastic} evolution of the
Ito type~\cite{gisin},
known previously for open statistical systems.
The importance of this formulation is
that it leads, under some rather generic assumptions,
to {\it localization} of the
quantum system in one of its states~\cite{gisin}.
This localization
occurs {\it independently} of the presence of a measuring
apparatus, and may be viewed as a state-vector
representation of the environment-induced decoherence
previously described in the
density matrix approach~\cite{ehns,emohn}.
According to the theorem of ref.~\cite{gisin},
the passage to classicality - in the above sense - will occur for
{\it all} non-isolated quantum systems, i.e., for all
realistic physical systems in our world. However, this
approach does not deal with the accuracy (uncertainty)
in the measurement. It is important to note, as we shall discuss
later on, that localization in this state-vector approach
does {\it not necessarily} imply
that the uncertainy in, say, length measurements,
will be reduced as a result of
decoherence. On the contrary, the opposite seems true.
Indeed, it has been known for some time
that decoherence effects
may lead to uncertainties in length measurements
that depend on the size of the distances measured~\cite{karo}.
\pr
It seems natural, in the light of the above discussion,
to expect that quantum gravity, in solving the apparent
incompatibility between quantum mechanics and classical gravity,
leads to a mechanism for decoherence~\cite{emohn,ehns,penrose}.
This can be realized by thinking of
quantum gravity as providing a space-time `medium'
through which low-energy quantum-mechanical particles
propagate~\cite{emn}.
The suggestion that quantum gravity induces decoherence
finds encouragement in some analyses of thought experiments,
in which the only assumption was that quantum gravity
should reproduce smoothly quantum mechanics and
classical gravity in their respective limits of validity.
These analyses indicate~\cite{karo,ng,gacmpla,outp9609}
that a length $L$ can only be defined operationally in
quantum gravity up to an uncertainty
$min \{ \delta L \}$
that is $L$-dependent, with the largest contribution
to this uncertainty~\cite{gacmpla,outp9609}
growing with $L$ like $\sqrt{L}$. Moreover, it is known~\cite{ng,karo}
that such an $L$-dependence in $min \{ \delta L \}$
can be associated with decoherence effects.
\pr
In recent years,
a candidate quantum-gravity model leading naturally
to decoherence has been provided by
a non-critical formulation of
string theory, exemplified by a treatment of quantum black
holes~\cite{emn}. At present, string is the only apparently
consistent quantum theory incorporating
gravity along with the rest of the fundamental interactions.
It should be noted that
non-critical strings constitute a more general framework than
critical ones, and that their precise quantum
structure is still not known. The approach of ref.~\cite{emn}
includes a treatment of the nature of time
in such non-critical strings.
It has been argued~\cite{emn} that time may be identified
with a renormalization-group scale on the world sheet
of the first-quantized version of the string. As such, the
passing of time is
associated with entropy production, reflecting information
loss across event horizons, microscopic as well as
macroscopic.
\pr
In this picture, space-time foam
is a medium of global, non-propagating gravitational
stringy degrees of freedom, which interact with propagating
low-energy particles via stringy gauge symmetries.
In the specific example of two-dimensional black hole studied
in ref.~\cite{emn}, these global Planckian modes fail to
decouple from the propagating low-energy modes in the presence
of microscopic black-hole space-time backgrounds.
Quantum coherence can be preserved only when the infinite set
of such modes are taken into account. This can be done in the
two-dimensional quantum field theory on the world sheet, but
local space-time
scattering experiments cannot detect these non-propagating
gravitational modes. Thus, the low-energy world  is effectively
an {\it open} system, and decoherence arises
in a stochastic way for specifically {\it stringy} reasons~\cite{emn}.
This is the first explicit realization of the idea that
the peculiar features of quantum gravity
may provide a `medium' for the propagation of ordinary matter.
Attempts have been made
to use these two-dimensional stringy black-hole models
as building blocks in the construction of
realistic four-dimensional quantum-gravity backgrounds
in string theory~\cite{emn,dbrane}. This is still an open issue, though,
and should not be considered as complete at this stage.
For our purposes, however, we shall view the two-dimensional
non-critical string~\cite{emn} as a concrete example of a
stochastic formulation of quantum gravity.

\pr
In this paper we discuss issues related to uncertainties
in measurement in the presence of quantum gravity entanglement
of the stochastic type mentioned above. We shall derive
bounds on the precision with which
lengths can be measured, that are in agreement with the general
arguments concerning quantum-gravity-induced decoherence
which were mentioned above. The structure of the article is as follows:
in section~2  we review the results of refs.~\cite{gacmpla,outp9609}
on the growth of the measurement error $min \{ \delta L \}$
with $L$ like $\sqrt{L}$. Then, in section~3
we review the type of vacuum entanglement
present in the non-critical string theory of ref.~\cite{emn},
discuss from various points of view its maximum possible
order of magnitude,
and motivate the ensuing modification of the state-vector
formalism of conventional quantum mechanics.
As we discuss in more detail in section~4, this suggests that
low-mass particle waves may experience dispersion as they
propagate through the space-time foam. If this has the
maximal order of magnitude discussed earlier, we show in
section~5 that it
leads naturally to $min \{ \delta L \} \sim \sqrt{L}$ as
in the previous approach to combining quantum mechanics and
classical gravity.
\pr
This striking observation adds support to
both approaches. It is important for the generic approach
of ref.~\cite{outp9609}, because it provides
a candidate microscopic theory which leads to bounds on measurement
errors of similar order of magnitude to those
derived previously on the basis of heuristic
considerations on the low-energy
limit of quantum gravity. For the string framework~\cite{emn}, it is
important because the derivation of the measurement bounds
relies, as we shall see, on an as yet unverified assumption
on the {\it maximal} order of magnitude of quantum-gravity
effects. The compatibility of the bounds derived within
the two different approaches constitutes an additional
argument in favour of the correctness of
these order-of-magnitude estimates for low-energy physics.
Section~6 contains a summary of our results, presents
some conclusions, and compares the distance uncertainty
derived here with various astrophysical limits.

\section{Bounds on the Possible Accuracy of Distance
Measurements from Heuristic Considerations
in Quantum Gravity}

Difficulties in introducing local observables
in quantum (and even classical) gravity
originate from the fact that general coordinate invariance
washes away the previous specification of individual
points of the space-time manifold using a coordinate system.
This specification of individual points
can be regained by labelling them using
particles of matter\footnote{This is an idea which has been
discussed extensively in the quantum gravity literature.
Refs.~\cite{rovelli,marolf}
recently reviewed this subject, presenting various viewpoints.},
{\it i.e.}, introducing
a material reference system (MRS), but then one must
take
into account the dynamics of this matter and its back-reaction
on the space-time manifold~\cite{rovelli}.

\pr
The problem of measuring distances in such a conceptual
framework was discussed in ref.~\cite{outp9609}, and
bounds were derived on the accuracy with which
distances can be measured in
an abstract MRS composed of identical
bodies, defining the space points,
with some internal physical variables,
defining the time instants~\cite{rovelli}.
The measurement procedure considered in ref.~\cite{outp9609}
(see also ref.~\cite{gacmpla}) involves
the exchange of a ``light signal'', {\it i.e.},
a signal propagating at the speed
of light and so composed of massless particles, between
two of the bodies composing the MRS.
Using 40-year-old purely quantum-mechanical arguments~\cite{wign},
it was shown that the uncertainty in measuring the distance
between two such MRS bodies must satisfy the following inequality:
\begin{eqnarray}
\delta L \ge
\sqrt{ {L \over M}}
~,
\label{dminqm}
\end{eqnarray}
where $M$ is the common
mass of the identical bodies composing the MRS,
and we use natural units such that $\hbar \! = \! c \! = \! 1$.
The condition (\ref{dminqm}) follows from the fact that, as their mass
is increased, the bodies of the MRS behave more and more classically
with respect to the other systems (clocks, detectors, etc.)
participating in the measurement process.
The classical laboratory limit is reached~\cite{outp9609},
in complete consistency with
quantum mechanics\footnote{Conventional quantum mechanics
is a theory that predicts the results
of measurements of (quantum) observables by some
classical (infinitely-massive) apparatus.
In this non-gravitational quantum-mechanical framework
there are no measurability bounds:
any observable can be measured with
perfect accuracy, at the price of renouncing all information
on its conjugate observable. The effects we discuss in this paper
go beyond this traditional uncertainty principle.},
in the $M \rightarrow \infty$ limit, in which case it is possible to
have $\delta L \sim 0$.

\pr
However, as discussed in ref.~\cite{gacmpla,outp9609},
the ideal scenario $\delta L \sim 0$
cannot be realized in presence of gravitational interactions, since
the required large values of the mass $M$ necessarily lead to great
distortions of the geometry. Well before reaching
the $M \! \rightarrow \! \infty$ limit, the
measurement procedure can no longer be accomplished in the way
envisaged in conventional quantum mechanics. In particular,
the minimal requirement\cite{outp9609} that a
horizon should not form around the bodies constituting
the MRS leads to the condition
\begin{eqnarray}
M \leq {s \over L_P^2}
~,
\label{mhor}
\end{eqnarray}
where $s$ is the common size of the bodies.
Eqs.~(\ref{dminqm}) and (\ref{mhor})
combine to yield the following measurability bound:
\begin{eqnarray}
\delta L \ge \sqrt{L L_P^2 \over s}
~.
\label{dmin}
\end{eqnarray}
In particular, since $L_P \sim 10^{-33}$cm
is expected to be the minimum length attainable in
a single-scale theory of quantum gravity such as may be
provided by string theory, it is interesting to consider the special
case $s \sim L_P$. In this case, eq.~(\ref{dmin}) reduces to
\begin{eqnarray}
\delta L \geq  \sqrt{L L_P}
~.
\label{dlminsplank}
\end{eqnarray}
As we shall see shortly, a bound of the same order
may be obtained from entirely different considerations
within certain stochastic approaches to quantum gravity.

\section{Space-Time Foam Effects on the Propagation of Low-Energy
Particles}
\pr
In this section we shall take a different approach,
discussing the propagation of low-energy particles through
space-time foam in the light of
an analysis of possible coherence loss
in quantum gravity, which is supported by one
formulation of non-critical string theory~\cite{emn}.
\pr
As we mentioned briefly in the Introduction,
in this approach we
identify time with a Liouville field which serves as a
(local) renormalization-group scale for
non-critical strings, leading to a stochastic evolution
in time which causes initially-pure
quantum-mechanical states to become mixed, with accompanying
entropy production~\cite{emn}.
A concrete example of such evolution
at a microscopic level is provided by two-dimensional
target-space string theory in black-hole backgrounds~\cite{witt,emn}.
In this theory,
the only propagating degrees of freedom, to be associated with
observables in local scattering experiments,
are massless scalar fields. However,
due to the persistence of infinite-dimensional stringy $W_{\infty}$
gauge symmetries in target space, there are
non-trivial couplings between the propagating
degrees of freedom and the `environment' of
global, delocalized stringy states, which may be interpreted as
solitons and figure among the higher-level string states~\cite{emn}.
These couplings disappear in flat space-times,
and appear only in the presence of microscopic fluctuations
in the space-time background, represented in this case by
black holes. This phenomenon constitutes
an explicit realization of the idea that quantum gravity
generates a space-time foam `medium', through
which low-energy matter propagates with non-trivial couplings.
It is suggested that
there are observable consequences of this evolution,
which follow from the fact that low-energy dynamics will be inherently
{\it open}, due to its couplings
to quantum-gravitational degrees of freedom, which find
an explicit model representation. In this approach,
time itself is a measure of information loss/entropy
production due to the interaction with this
microscopic environment. As a result, an open-system
modification of conventional
quantum-mechanical time evolution, with
genuine $CPT$ violation~\cite{cplear}, will characterize the temporal
evolution of the low-energy (observable) world.
The stochastic nature of the quantized
renormalization-group flow on the
world sheet yields stochastic
time evolution according to this picture.
\pr
We provide below a brief review of the
corresponding modifications of low-energy
quantum mechanics believed to be induced by such an
approach to quantum gravity, specifically in the
context of the non-critical string analysis of ref.~\cite{emn}.
Our intention here is only to
recapitulate briefly the basic results of the approach
for the benefit of the non-workers in the field, and
to estimate the maximum possible order of magnitude of
deviations from conventional quantum mechanics. Further
details can be found in ref.~\cite{emn}.
\pr
The evolution of pure states into mixed ones necessitates
the introduction of density matrices $\rho (t)$
for the description of low-energy dynamics~\cite{ehns}.
The modified form of temporal evolution may be deduced~\cite{emn}
from the renormalization-group invariance
of physical quantities in non-critical
string theory, and may be written as a modified quantum
Liouville equation for $\rho(t)$~\cite{emn,kogan}:
\bea
&~&
  Q\frac{\partial}{\partial t} \rho \equiv Q {\dot \rho}
  = i [ \rho, H] + {\delta\H} \rho
\qquad ; \qquad \delta\H \equiv i Q{\dot g}^i G_{ij} [g^j,~ ]
\nn \\
&~& Q=\sqrt{\frac{|25-c[g,t]|}{3}}
\label{s1}
\ee
\nk where the $g^i$ are fields representing string couplings,
$G_{ij}$ is a metric on the space of these couplings,
and $c[g]$ is the running central charge of the
non-critical string. It is important to stress that
the first-quantized approach to non-critical strings entails
quantization of the background fields $g^i$,
as a consequence of resummation over higher-genus world-sheet
topologies in the Polyakov path integral over string histories.
The quantities $Q{\dot g}^i \equiv \beta ^i$ are the
world-sheet renormalization-group $\beta$-function coefficients
in a formulation of non-critical string theory in terms of
general two-dimensional field theories
($\sigma$ models) on the world sheet.
In critical string theory, the coefficients
$\beta ^i$ vanish, as a result of world-sheet
conformal invariance.
In the presence of fluctuations in the space-time
background, such as quantum black holes, however,
apparent conformal invariance is lost for a low-energy
observer who observes only propagating fields. From the explicit
example of two-dimensional (spherically-symmetric
four-dimensional) stringy black holes,
we know that conformal invariance is
guaranteed only if the infinite set of global string modes
is turned on as backgrounds. The latter cannot be observed
by localized scattering experiments. When these are not taken
into account, the result is that the effective
$\beta ^i$ do not vanish, inducing Liouville field- (scale-)
dependence of the renormalized coupling/fields $g^i$,
as described in ref.~\cite{emn}.
In a quantum theory of non-critical strings, such
apparent violations of world-sheet conformal invariance are allowed,
in the sense that one can dress up the theory
with Liouville scale factors so as to restore conformal symmetry.
It is this dressing that leads to the flow of
time in the interpretation of ref.~\cite{emn,kogan}, and it is
the semi-group nature of the renormalization group that causes
this time flow to have an arrow, accompanied by coherence loss
and entropy gain.
\pr
At present we are not in a position to calculate
reliably the order of magnitude of any
such deviation from conformal invariance. However,
various different methods point towards similar
estimates, as we now discuss.
\begin{itemize}
\item
In the {\it non-critical-string}
framework~\cite{emn}, within which the mechanism for the `opening'
phenomenon is traceable directly to Planckian physics,
we expect
that the string $\sigma$-model coordinates $g^i$ obey
renormalization-group equations of the general form
\be
    Q{\dot g}^i = \beta ^i M_{P} \qquad : \qquad
    |\beta ^i | = {\cal O}\left(\frac{E^2}{M_{P}^2}\right)
\label{s3}
\ee
where the dot denotes differentiation
with respect to the target time, measured in string $(L_P=M_{P}^{-1})$
units, and $E$ is a typical energy scale in the low-energy
observable matter system. Since the metric $G_{ij}$ and
the couplings $g^i$ are themselves dimensionless
numbers of order unity, we expect that
\be
|\delta\H | = {\cal O}\left(\frac{E^2}{M_{P}}\right)
\label{s4}
\ee
in general. This is a {\it maximal} estimate for such effects.
Formally, (\ref{s3}) stems from a (target-space) derivative
expansion of the closed-string $\beta^i$, of the generic form
$\sum _{n=1}^{\infty} c_n (\alpha ' \partial _{X^M}^2)^n$,
where we keep only the first term. However, the
coefficient $c_1$ is not at present
known for any physical system.
There are expected to be system-dependent numerical factors
that depend on the underlying string model,
$|\delta\H |$ might be suppressed by further
($E/M_{P}$)-dependent factors, and it could even vanish for
some systems. Nevertheless, equation (\ref{s4})
gives us a  natural order-of-magnitude estimate
for the decoherence effects due to stringy quantum gravity
in a generic physical system.
\item
More heuristic {\it quantum gravity} considerations
yield a similar order-of-magnitude estimate. The term
proportional to $\delta\H$ in (\ref{s1}) can be interpreted
as a non-canonical contribution to the rate of change of
the density matrix $\rho$ due, in some sense, to `scattering'
off evanescent black holes~\cite{ehns}, which one might be able to
estimate by the generic formula
\be
\delta (Q \dot \rho) \simeq \sigma {\cal N} \rho
\label{emon}
\ee
where $\sigma$ is the `scattering cross section', and
$\cal N$ is the number density of microscopic
`target' black holes. Within this heuristic approach, it
is natural to expect that $\sigma$ would be proportional
to the square of some `scattering amplitude' obtainable from some
effective particle/black hole interaction. The latter could be
expected to contain a factor ${\cal O}(1/M_P^2)$, as for example
in a four-fermion interaction of the generic form~\cite{ehns,emohn}
\be
{\cal L} = {\cal O}({1 \over M_P^2}){\bar \psi \psi} {\bar \psi \psi}
a_{BH} \, + \, h.c.
\label{worm}
\ee
where the $\psi$ are fermionic fields,
$a_i$ is a black hole annihilation operator, and the
corresponding creation operator appears in the Hermitian
conjugate term in (\ref{worm}). Squaring the generic
`scattering amplitude' generated by the interaction
(\ref{worm}) yields a factor ${\cal O}(1/M_P^4)$ in the
cross section $\sigma$ in (\ref{emon}). We would expect
that the black hole density
\be
{\cal N}={\cal O}( M_P^3 )
\label{density}
\ee
corresponding to ${\cal O}(1)$ microscopic Planck-mass
black hole per Planck volume in space, in much the same
way as one expects one QCD instanton per unit volume
$ \simeq \Lambda_{QCD}^3$ in the strong-interaction vacuum.
We know of no evidence from generic quantum-gravity
considerations, or more specifically from string theory,
which would argue for any parametric suppression of the
generic estimate (\ref{density}), e.g., of the form
$e^{-1/\alpha}$ or $e^{-1/\sqrt\alpha}$. Combining
(\ref{worm}) and (\ref{density}), therefore, we arrive~\cite{emohn}
at the estimate
\be
\delta (Q \dot \rho) ={\cal O}( {E^2 \over M_P})
\label{second}
\ee
where the energy factors in the numerator follow from naive
dimensional analysis.
\item
The above estimates receive some support from a
{\it semi-classical} black-hole
calculation, in which a scalar field with gauge
interactions is quantized in
a four-dimensional black-hole background.
Calculating the dependence of the density matrix for the
external scalar field on the radius of the
black-hole horizon, and recalling
that the latter evolves with
time as a result of Hawking radiation, we have
arrived~\cite{winstan} at the same order-of-magnitude estimate
for the `open' term in the modified quantum Liouville
equation (\ref{s1}).
\end{itemize}
\pr
The next step is to explore the consequences of
this estimate for the propagation of what would
normally be a coherent quantum-mechanical wave.
Although we have used above the density-matrix formalism,
which is in many applications the most appropriate
for a description of mixed states,
it is also possible to cast the
problem of environmental entanglement
in a state-vector formalism, and this turns out to be
instructive for a first discussion of wave propagation.
We follow the approach due to
Gisin and Percival~\cite{gisin},
which involves stochasticity. As we argued above,
stochasticity is an essential feature of
the identification of target time with a (stochastic)
renormalization group scale variable on the world sheet.
\pr
The Lindblad formalism~\cite{lindblad,ehns}
admits a state-vector picture involving a
modified Schr\"odinger equation~\cite{gisin}.
If $|\Psi >$ is the state vector,
environmental entanglement may be represented as
a {\it stochastic} differential Ito process
for $|\Psi >$~\cite{gisin}:
\bea
&~&Q|d\Psi > =-\frac{i}{\hbar} H |\Psi > dt +
\sum _m (<B^\dagger _m >_{\Psi} B_m - \frac{1}{2}
B^\dagger _m B_m - \nn \\
&~&\frac{1}{2} <B^\dagger _m>_{\Psi}
<B_m>_{\Psi })~|\Psi > dt +
\sum _m
(B_m - <B_m>_{\Psi})~|\Psi > d\xi _m
\label{s8})
\eea
where $H$ is the Hamiltonian of the system,
and $B$, $B^\dagger$ are `environment' operators.
In our case, these may be
defined as appropriate `squared roots'
of the various partitions of the operator
$\beta ^j G_{ij} \dots g^i $~\cite{emn},
and the $d\xi _m$ are complex differential
random variables, associated with stochastic
white-noise Wiener or Brownian processes, which
represent environmental effects averaged out by a low-energy
local observer. In our quantum-gravity/string-theory picture,
such effects are due to global Planckian modes~\cite{emn}
of the string which cannot be detected by local scattering
experiments. As such, these
effects are therefore inherently stringy/quantum
gravitational.
\pr
It is clear from (\ref{s8}) that among
the effects of the environmental/quantum-gravity entanglement
represented by the operators $B$ are
extra, frequency-dependent
phase shifts in the Schr\"odinger wavefunction for the modes $g^i$,
defined by $<g^i|\Psi> = \Psi _{g^i}$,
as well as attenuation effects on the amplitudes of the waves.
All such effects are expected to be of the
generic order of magnitude
$O[B^\dagger B] = O[\beta^i \dots G_{ij}g^j] $.
The frequency-dependent attenuation effects
can be shown to lead to decoherence, in the sense
of an exponential (with time)
vanishing of the off-diagonal elements
of the density matrix~\cite{emohn} in
(\ref{s1}). In the case of interest here,
the precise order of such effects can be found in principle
by solving (\ref{s8}) in the case of massless modes
propagating in a stringy quantum gravity medium,
as appropriate for the {\it gedanken} experiments of section 2.
\pr
\section{Wave Dispersion in Non-Critical String Backgrounds}
\pr
A more detailed discussion of the possible dispersive
effects of space-time foam on wave propagation requires
a deeper discussion of the non-critical string approach~\cite{emn}
than we have provided so far in this paper. We will now
recall some relevant features of this formalism, and then
explore their specific consequences for wave dispersion.
\pr
An appropriate quantum formalism for discussing
non-critical string is that of generalized
$\sigma$ models on a fixed lowest-genus topology of the world sheet,
with dressing by a Liouville field $\phi$, which in the interpretation
of ref.~\cite{aben,emn,kogan} plays the r\^ole of target time.
The relevant formalism is that of non-conformal
deformations $V_i$ of $\sigma$-model actions  $S[g]$:
\be
   S[g] = S[g^*] +
   \int d^2 z \sqrt{\gamma} g^i V_i
\label{C1}
\ee
where the $\{ g^i \}$ denote massless
background fields in the string target space,
$\{ g^* \}$ represents an equilibrium conformal
solution around which we perturb, and
$\gamma$ is the world-sheet metric.
In the approach of ref.~\cite{emn},
for generic curved world sheets the renormalization
scale may be taken as
a local function on the Riemann surface,
which then is quantized to incorporate
fluctuating topologies. This necessitates
the introduction of the Liouville mode $\phi$,
which in this approach plays the r\^ole
of a covariant quantum renormalization scale~\cite{emn,kogan}:
$\gamma = e^{\phi} \gamma^*$, where $\gamma^*$ is a reference
world-sheet metric.
The renormalization-group picture we adopt is that of
Wilson, according to which non-trivial scaling
follows from integrating out degrees of freedom in
an effective theory consisting of the massless modes
$\{ g^i \}$ only.
\pr
Coupling the theory (\ref{C1}) to two-dimensional quantum gravity
restores conformal invariance at the quantum level,
by making the gravitationally-dressed operators
$[V_i]_{\phi}$ {\it exactly} marginal, {\it i.e.}, ensuring
the absence of any covariant dependence on a
world-sheet scale. The final result for the gravitationally-dressed
matter theory is then
\be
S_{L-m} = S[g^*] + \frac{1}{4\pi\alpha '}
\int d^2 z \{\partial _\alpha \phi \partial ^\alpha \phi
- QR^{(2)} + \lambda ^i(\phi ) V_i \}
\label{C5}
\ee
where $\alpha '$ is the inverse of the string tension.
To order $O(g^2)$, the gravitationally-dressed
couplings $\lambda (\phi )$ are given by:
\be
\lambda ^i(\phi ) =g^i e^{\alpha _i \phi }
+ \frac{\pi}{Q \pm 2\alpha _i } c^i_{jk} g^jg^k
~\phi~e^{\alpha _i \phi } + \dots
\label{C6}
\ee
with
\be
\alpha _i ^2 + \alpha _i Q =sgn(25-c)(h_i - 2)
\label{C7}
\ee
where the $h_i -2$ are the scaling dimensions
of the operators $V_i$ before Liouville dressing.
The operator product expansion coefficients $c^i_{jk}$
are defined as usual by coincident limits of the products
of pairs of vertex operators $V_i$:
\be
lim_{\sigma \rightarrow 0} V_j (\sigma )V_k (0) \simeq
c^i_{jk} V_i (\frac{\sigma}{2}) + \dots
\label{C3}
\ee
where the completeness of the set $\{ V_i \}$ is assumed.
\pr
~From the quadratic equation (\ref{C7}) for $\alpha _i$,
only the solution
\be
\alpha _i = -\frac{Q}{2} +
\sqrt{\frac{Q^2}{4} - (h_i - 2)}
\label{solutions}
\ee
for $c \ge 25 $ is kept, due to the Liouville
boundary conditions. It is worth noticing that
the case $c \ge 25$ automatically implies a Minkowski signature
for the Liouville mode $\phi $, enabling it to be
interpreted as target time~\cite{emn,aben}.
Indeed, equation (\ref{s1}) was derived~\cite{emn}
by identifying
the physical Minkowski time $t$ with $- i \phi$,
an identification which has been supported by model black-hole
calculations.
\pr
The structure of eq.~(\ref{C6}) suggests that the
effects of the non-criticality are quite complicated
in general. However, the form of the Liouville time dependence
implies that one of the physical effects of the
non-criticality is a modification
of the time-dependence of the $\lambda ^i$, which may be
described to order $O(g^2)$ by the following approximation
to (\ref{C6}):
\be
   \lambda ^i (t ) \sim g^i e^{ i(\alpha _i + \Delta \alpha _i)t}
\label{estimate}
\ee
where the $\Delta \alpha _i $ depend on the $c^i_{jk}$,
which encode the interactions with quantum-gravity
fluctuations in the space-time background.
\pr
We are now ready to discuss the issue of wave dispersion.
~From a target-space point of view, the $g^i$ may be viewed
as the Fourier transforms, {\it i.e.}, the
polarization tensors, of target-space background fields,
and the vertex operators are wave
operators~\footnote{Here the concepts of Fourier
transforms and plane waves should be understood
as being appropriately generalized to curved target spaces,
with the appropriate geodesic distances taken into account.
For our purposes below, the details of this will not be relevant.
We shall work with macroscopically-flat space-times, where
the quantum-gravity structure appears
through quantum fluctuations of the vacuum, leading
simply to non-criticality of the string, in the sense
of non-vanishing $\beta ^i$ functions.}.
For instance, for a deformation corresponding to a
scalar mode $T(X)$, one can write
\be
   \int d^2 z \sqrt{\gamma} g^i V_i \equiv
\int d^2 z \sqrt{\gamma} \int d^D k e^{i k_M X^M(z,{\bar z})}
{\tilde T}(k)
\label{scalar}
\ee
where the summation over $i$ includes target-space integration
over $k$, on a $D$-dimensional Euclidean space.
For massless string modes, as opposed to higher modes,
$h_i-2 = k^Mk_M $.
For our purposes, it suffices to consider the case of an
almost flat space time with small quantum-gravity corrections
coming from the interactions of massless low-energy modes
with the environment of Planckian string states,
implying that the string is close to a fixed point
for which $Q^2 = c[g^*]-25 = 0$. This is the case in certain
cosmological backgrounds of interest to us~\cite{tseytl}.
In such a case, we see from (\ref{solutions}) that
$\alpha _i \sim |\vec{k}|$.
\pr
Our basic working hypothesis, which has been confirmed
explicitly in the two-dimensional string black hole
example~\cite{emn},
is that the matter deformations in such a black-hole background
are not exactly marginal unless the couplings to
non-propagating Planckian global modes
are included in the analysis. This implies that the set of the
operator product expansion coefficients
$c^i_{jk}$ in (\ref{C6}) includes non-zero couplings between
massless and Planckian modes in the presence of non-trivial
metric fluctuations of black-hole type, and more generally
in the presence of structures with a space-time
boundary. These couplings arise
in string theory from the infinite set of
stringy gauge $W_\infty$ symmetries mentioned in the previous
section. They imply that the massless modes constitute an open
system, and that they lose information to these non-propagating
higher-level modes which are not detected
in local scattering experiments, inducing apparent
decoherence~\cite{emn}.
\pr
The order of magnitude of such couplings is
not known at present, and precise calculations would require
a full second-quantized string theory. However,
in the same spirit of the {\it maximal} estimate (\ref{s4}),
which was also adopted in ref.~\cite{emn,cplear},
one may assume that
\be
\Delta \alpha _i \sim \eta {|\vec{k}|^2 \over M_P}
\label{etadef}
\ee
with $\eta$ a dimensionless quantity, which
parametrizes our present
ignorance of the quantum structure of space time.
\pr
For the purposes of the present work,
we restrict ourselves to non-critical strings on
fixed-genus world sheets, in which case $\eta$ in (\ref{etadef})
is real. This should be viewed as describing only part of the
quantum-gravity entanglement, namely that associated with the
presence of global string modes~\cite{emn}. The full
string problem
involves a summation over genera, which in turn implies
{\it complex} $\eta$'s, arising from the appearance
of imaginary parts in Liouville-string correlation functions
on resummed world sheets, as a result of instabilities pertaining to
microscopic black-hole decay in the quantum-gravity
space-time foam~\cite{emn}. Such imaginary parts
in $\eta$ will produce frequency-dependent
attenuation effects in the amplitudes of quantum-mechanical
waves for low-energy string modes. This type
of attentuation effect leads to
decoherence of the type appearing in the density-matrix
approach to measurement theory~\cite{emohn,emn}.
For our purpose of deriving bounds on the possible
accuracy of distance measurements, however,
such attenuation effects need not be taken into account, and
the simple entanglement formula (\ref{etadef}), with real $\eta$,
will be sufficient.
\pr
The value of $\eta$ depends in general on the type of massless
field considered. In particular, in the case of photons
$\eta$ is further constrained
by target-space gauge invariance, which restricts
the structure of the relevant $\sigma$-model $\beta$ function.
With this caveat in mind, (\ref{etadef}) represents a maximal
estimate of $\Delta \alpha _i$, compatible
with the generic structure of perturbations of the $\sigma$-model
$\beta$ functions for closed strings.
\pr
The equations (\ref{C6}), (\ref{scalar}) and (\ref{etadef})
indicate that the dressed $\sigma$-model deformation (\ref{C5})
corresponds to waves of the form\footnote{Note that eq.~(\ref{wave})
is consistent with target-space gauge
invariance, since it may describe
one of the polarization states of the photon, and
the transversality condition on the photon
polarization tensor, $k^\mu A_\mu = 0$, can be
maintained.}
\be
 e^{i |{\vec k}| t + i{\vec k}. {\vec X}  +
 i \eta {|\vec{k}|^2 \over M_P} t }
\label{wave}
\ee
The corresponding modified dispersion relation
\be
E \! \sim \! |\vec{k}| + \eta |\vec{k}|^2 /M_P,
\label{dispersion}
\ee
which was dictated by the overall conformal invariance
of the dressed theory,
implies that massless particles propagate in the quantum-gravity
`medium' with a velocity that is effectively
energy-dependent:
$v \! \sim \! 1+ \eta E/M_P$.
\pr
\section{Distance Measurement Errors within the
Non-Critical String Approach}
\pr
We now use the result (\ref{wave})
to derive bounds on the precision with which macroscopic distances
can be measured, from the quantum-gravity point of view
suggested by our non-critical approach to string theory.
For comparison with the previous estimate of section 2,
we consider the use of massless probes
propagating through space-time foam, modelling the latter
as a stochastic environment. We consider only
contributions to the distance uncertainty that are associated with
the emission and propagation of such
probes. All other possible
contributions to the uncertainty, associated, for example, with
the finite extent of the `clock' used to measure time delays,
will be ignored.
\pr
We start by considering the contribution to the distance
uncertainty associated with the vacuum entanglement of the probe
during its motion between the bodies whose distance
is to be measured. The fact that (\ref{wave}) describes
particle propagation with an energy-dependent speed $v(E)$,
leads to a (positive or negative, depending on the sign of $\eta$)
time shift, with respect to conventional relativistic
$v \! = \!1$ propagation. After a time $t$, this shift amounts to
\be
\Delta t \sim \eta \, L_P \, E \, t
~.
\label{td}
\ee
~From the point of view of an observer attempting to use the time of
travel $T$ of one such probe to measure a length $L$, this correction
can be taken systematically into account. However, an
uncertainty in the relation between $T$ and $L$
is introduced by the quantum uncertainty
in the energy of the probe. Specifically, the time of travel
of a wave packet of
energy spread $\delta E$ exchanged between two bodies whose distance
$L$ is being measured, has an uncertainty
\be
\delta L_1 \sim \eta \, L_P
\, \delta E \, T \sim
\eta \, L_P  \, \delta E \, L
~,
\label{dla}
\ee
where we have used on the right-hand side the fact that the time of travel
is, for a typical experimental set up, of order $L$.
We also keep track of the fundamental
uncertainty for a low-energy observer, in quantum
gravity~\cite{padma} and
string
theories~\cite{venezkonish}, due to the natural cut-off in lengths
set by the Planck or string length:
\be
\delta L_2 \sim L_P
~.
\label{s13c}
\ee
In the measurement analysis, these contributions
of quantum-gravitational origin must be combined
with the other, well known, purely quantum-mechanical contributions
to the uncertainty. In particular, it follows
from the time-energy uncertainty principle that
\be
\delta E \delta T^{emission} \sim 1
\label{s15b}
\ee
where $\delta T^{emission}$ is the uncertainty in the time of emission
of the probe. Since this obviously contributes
to the uncertainty in the time of travel measured in the experiment,
from (\ref{s15b}) one finds the contribution
\be
\delta L_3 \sim \frac{1}{\delta E}
\label{s16}
\ee
to the uncertainty in the measurement of $L$.
\pr
$~$From (\ref{dla}), (\ref{s13c}), and (\ref{s16})
one finds a total uncertainty in the length measurement
that may be written as
\be
\delta L_{total} \ge \eta L_P L \delta E + \frac{1}{\delta E}
+ L_{P}
\label{s15}
\ee
The observer can minimize the uncertainty (\ref{s15})
only by tuning $\delta E$, {\it i.e.}, by preparing a
suitable wave packet. It is, then, straightforward to see that
the following bound holds for the error in the measurement of $L$:
\be
min \{ \delta L \} \sim  \sqrt {\eta L L_{P}} + L_P
~.
\label{s17}
\ee
in the context of the non-critical string approach to quantum
gravity outlined in the previous section.
\pr
If, possibly as a result of string-theoretical cancellations
as yet unknown, the actual wave dispersion due to space-time
foam turned out to be weaker
than the maximal order-of-magnitude estimate (\ref{s4}),
a correspondingly weaker dependence on $L$ of $min \{ \delta L \}$
would be found. For example, as the reader can easily verify,
if (\ref{s4}) is
replaced by $|\delta\H | = {\cal O}(\frac{E^n}{M^{n-1}_{P}})$,
one finds that $min \{ \delta L \} \sim L^{1/n} \, L_P^{1-1/n}$.

\section{Summary and Conclusions}
\pr
We have derived bounds on the precision of macroscopic
distance measurements which follow from
a stochastic approach to quantum gravity,
motivated by the formulation of
non-critical string theory in ref.~\cite{emn}.
Within this approach, distance uncertainties arise
from the observation that propagating particles
are subject to dispersion by the foamy `medium'
of microscopic fluctuations in space time, causing them to
acquire an energy- and time- dependent phase shift.
This is a result of their
entanglement with the quantum-gravitational `environment', and
such dispersion would lead to
an $L$-dependent limit on the accuracy with which a
macroscopic length $L$ could be measured.
\pr
We have reviewed and developed previous arguments suggesting
a specific maximal estimate (\ref{s4}) for the
order of magnitude of the quantum-gravitational entanglement,
which leads to a limit on the accuracy of distance
measurement that is proportional to $\sqrt{L}$.
Interestingly, this is of the same order as the bound
predicted by a previous independent analysis combining
quantum mechanics and classical general
relativity~\cite{gacmpla,outp9609},
which we briefly reviewed in Sec.2. We find this similarity
remarkable, because the two physical
mechanisms involved in deriving the $\sqrt{L}$-dependent
bound are {\it a priori} different and independent.
This agreement could even be construed as
providing additional support for the assumption
(\ref{s4}).
\pr
It is important to note
that the type of vacuum entanglement
effects
appearing in (\ref{s4}) are within the range allowed
by
the
present
experimental accuracy.
For instance,
for lengths of the order of the size of the present Universe,
$10^{10}$ light years, the formula (\ref{s17}) predicts
a distance uncertainty of only $\eta^{1 \over 2} \, 10^{-3}~cm$.
Even the most accurate astrophysical experimental data
appear to constrain only marginally the value of our
phenomenological parameter $\eta$.
Pulsars provide an excellent long-distance laboratory,
because of their excellent timing. However,
even for $1$ GeV signals, at the upper end of the energies
of observed pulsar signals, and for a pulsar distance
of the order of $10^{4}$ light years, which would correspond
to a pulsars on the other side of our galaxy,
the energy-dependent time delay described in (\ref{td})
is only of order $\eta \, 10^{-8} s$.
Similarly, (\ref{td}) predicts that the neutrinos from
supernova 1987a,
which had energies of order $10$ to $100$ MeV
and came from a distance of about $1.6 \, 10^{5}$ light years,
might have experienced a
time dispersion of only about $\eta \, 10^{-7} s$
as a result of their entanglement with the quantum-gravity
medium/vacuum.

\pr
This brief discussion indicates that a value $\eta \sim 1$,
corresponding to the maximal estimate (\ref{s4}),
may not be in contradiction with any present data. However,
a more systematic investigation of terrestrial and
astrophysical data is needed
to set an accurate experimental bound on $\eta$.
To the extent that the maximal estimate (\ref{s4}) cannot
be excluded, experimental searches for
quantum-gravity effects in particle and other laboratory
physics motivated by this estimate should not be discouraged.
We would cite in particular the searches
for open quantum-mechanical $CPT$ violation
in neutral kaons\cite{cplear}, which could even be
on the verge of experimental observation, if (\ref{s4}) holds,
and also possible searches in macroscopic quantum-mechanical
systems such as SQUIDs~\cite{emohn,squid}.

\bigskip
\bigskip
\bigskip
\nk {\Large {\bf Acknowledgements} }

It is a pleasure to acknowledge helpful discussions on astrophysics
and cosmology with S. Sarkar. Two of us (N.E.M. and D.V.N.)
thank P. Pavlopoulos and the CPLEAR collaboration for their
interst and support during part of this work.
G.A.-C. acknowledges financial support from the
European Union under contract \#ERBCHBGCT940685.

\newpage
\baselineskip 12pt plus .5pt minus .5pt

\end{document}